 \documentclass[final,3p,times,twocolumn]{elsarticle}
\usepackage{amssymb}
\biboptions{sort&compress}

\journal{NIM B373(2016)17}

\begin{document}

\begin{frontmatter}



\title{Activation cross sections of deuteron induced reactions on niobium in the 30-50 MeV energy range}


\author[1]{F. Ditr\'oi\corref{*}}
\author[1]{F. T\'ark\'anyi}
\author[1]{S. Tak\'acs}
\author[2]{A. Hermanne}
\author[3]{A.V. Ignatyuk}
\cortext[*]{Corresponding author: ditroi@atomki.hu}

\address[1]{Institute for Nuclear Research, Hungarian Academy of Sciences (ATOMKI),  Debrecen, Hungary}
\address[2]{Cyclotron Laboratory, Vrije Universiteit Brussel (VUB), Brussels, Belgium}
%\address[3]{Cyclotron Radioisotope Center (CYRIC), Tohoku University, Sendai, Japan}
%\address[4]{Nuclear Research Center – Egyptian Atomic Energy Authority, Cairo, Egypt}
\address[3]{Institute of Physics and Power Engineering (IPPE), Obninsk, Russia}

\begin{abstract}
Activation cross-sections of deuterons induced reactions on Nb targets were determined with the aim of different applications and comparison with theoretical models. We present  the experimental excitation functions of $^{93}$Nb(d,x)$^{93m,90}$Mo, $^{92m,91m,90}$Nb,$^{89,88}$Zr and $^{88,87m,87g}$Y in the energy range of 30–50 MeV. The results were compared with earlier measurements and with the cross-sections calculated by means of the theoretical model codes ALICE-D, EMPIRE-D and TALYS (on-line TENDL-2014 and TENDL-2015 libraries). Possible applications of the radioisotopes are discussed in detail.
\end{abstract}


\end{frontmatter}

% \linenumbers

\section{Introduction}
\label{1}
Activation cross sections of deuteron induced reaction on niobium are important for several practical applications as well as for basic nuclear physics needed for verification and improvement of nuclear reaction models. In the frame of a systematic study of the light ion induced reactions on structural materials, we reported earlier experimental cross section data  on niobium for protons up to 67 MeV \cite{1, 2}, deuterons up to 38 MeV \cite{3, 4} and alpha particles up to 43 MeV \cite{5} and discussed the possible applications in more detail. As no datasets above 40 MeV for deuteron induced reactions are available (up to 40 MeV only our earlier results and a recent set of measurements by Avrigeanu \cite{6}, we extended the energy range of the experimental data up to 50 MeV in this work and  included a comparison with theoretical calculations using different model codes.

\section{Experiment}
\label{2}
For the cross section determination an activation method based on stacked foil irradiation followed by $\gamma$-ray spectroscopy was used. The stack consisted of a sequence of Rh(26 $\mu$m), Al(50 $\mu$m), Al(6 $\mu$m), In(5$\mu$m), Al(50 $\mu$m), Pd(8 $\mu$m), Al(50 $\mu$m), Nb(10 $\mu$m), Al(50 $\mu$m) foils, repeated 9 times and bombarded for 3600 s with a 50 MeV deuteron beam of nominal 32 nA at Louvain la Neuve (LLN) Cyclotron Laboratory. The beam current was more exactly estimated in the Farady cup and corrected by using the monitor reactions.
The activity produced in the targets and monitor foils was measured non-destructively (without chemical separation) using high resolution HPGe gamma-ray spectrometers (made by Canberra, coupled with a Multichannel analyzer running with the Genie 2000 software ©). Three series of $\gamma$-spectra measurements were performed starting at 9.1-10.1h, 28.6-44.7 h, and 458.4-533.4 h after EOB, respectively.
The evaluation of the gamma-ray spectra was made by both a commercial \cite{7} and an interactive peak fitting code \cite{8}.
The cross-sections were calculated by using the well-known activation formula with measured activity, particle flux and number of target nuclei as input parameters. Some of the radionuclides formed are the result of cumulative processes (decay of metastable states or parent nuclides contribute to the formation process). Naturally occurring niobium is monoisotopic ($^{93}$Nb) and hence $^{93}$Nb(d,x) reaction cross-sections (direct formation or cumulative production) are presented.
The decay data were taken from the online database NuDat2  \cite{9} and the Q-values of the contributing reactions from the Q-value calculator \cite{10}, both are presented in Table 1.
Effective beam energy and the energy scale were determined initially by a stopping calculation \cite{11} based on estimated  incident energy and target thickness and finally corrected  \cite{12} on the basis  of the excitation functions of the  $^{24}$Al(d,x)$^{22,24}$Na  monitor reactions \cite{13}  simultaneously re-measured over the whole energy range. For estimation of the uncertainty of the median energy in the target samples and in the monitor foils, the cumulative errors influencing the calculated energy (incident proton energy, thickness of the foils, beam straggling) were taken into account. The uncertainty on the energy is in the $\pm$ 0 .5 - 1.5 MeV range, increasing towards the end of stack. The individual uncertainties occurred in the propagated error calculation are: absolute abundance of the used $\gamma$-rays (5\%), determination of the peak areas (4-10\%), the number of target nuclei (beam current)(5\%), detector efficiency (10\%). The total uncertainty of the cross-section values was estimated at 10–15\%.
The beam intensity (the number of the incident particles per unit time) was obtained preliminary through measuring the charge collected in a short Faraday cup and corrected on the basis of the excitation functions of the monitor reactions compared to the latest version of IAEA-TECDOC-1211 recommended data base \cite{13}.
The uncertainty on each cross-section was estimated in the standard way by taking the square root of the sum in quadrature of all individual contributions, supposing equal sensitivities for the different parameters appearing in the formula. The following individual uncertainties are included in the propagated error calculation: absolute abundance of the used $\gamma$-rays (4–11\%), determination of the peak areas including statistical errors (5\%), the number of target nuclei including non-uniformity (5\%), detector efficiency (10\%) and incident particle intensity (7\%).The total uncertainty of the cross-section values was evaluated to approximately 8–14\%. \cite{14}.

\begin{table*}[t]
\tiny
\caption{Decay and nuclear characteristic of the investigated reaction products, contributing reactions and their Q-values}
\centering
\begin{center}
\begin{tabular}{|p{0.7in}|p{0.5in}|p{0.7in}|p{0.5in}|p{0.5in}|p{0.8in}|p{0.6in}|} \hline 
\textbf{Nuclide\newline Spin/parity\newline Isomeric level} & \textbf{Half-life} & \textbf{Decay method\newline } & \textbf{E$_{\gamma}$(keV)} & \textbf{I$_{\gamma}$(\%)} & \textbf{Contributing process} & \textbf{Q-value\newline (keV)} \\ \hline 
${}^{93}$${}^{m}$Mo\newline 21/2$^{+ }$\newline 2424.97 keV & 6.85 h & IT 100\% & 263.049 684.693\newline 1477.138 & 57.4 \newline 99.9\newline 99.1 & $^{93}$Nb(d,2n) & -3413.59 \\ \hline 
${}^{90}$Mo\newline 0${}^{+}$ & 5.67 h & EC 75.1 \%\newline ${\beta}^{+}$ 24.9 \% & 122.370\newline 162.93\newline 203.13\newline 257.34\newline 323.20\newline 445.37\newline 941.5\newline 1271.3 & 64\newline 6.0\newline 6.4\newline 78\newline 6.3\newline 6.0\newline 5.5\newline 4.1 & $^{93}$Nb(d,5n)\newline  & -34260.98 \\ \hline 
$^{92m}$Nb\newline 2$^{+}$\newline 135.54keV & 10.15 d & EC 99.935 \%\newline ${\beta}^{+}$ 0.065 \% & 912.6\newline 934.44 & 1.78\newline 99.15 & $^{93}$Nb(d,p2n)\newline  & -11055.13 \\ \hline 
$^{91m}$Nb\newline 1/2$^{-}$\newline 104.605 keV & 60.86 d & EC 3.4 \%\newline IT 96.6\% & 104.62\newline 1204.67 & 0.574\newline 2.0 & $^{93}$Nb(d,p3n) & -18941.62 \\ \hline 
$^{90}$Nb\newline 8+ & 14.60 h & EC 48.8 \%\newline ${\beta}^{+}$ 51.2 \% & 132.716\newline 141.178 & 4.13\newline 66.8 & $^{93}$Nb(d,p4n)\newline $^{90}$Mo decay\newline  & -30989.3\newline -34260.98 \\ \hline 
${}^{89}$Zr\newline 9/2$^{+}$ & 78.41 h & EC 77.6 \%\newline ${\beta}^{+}$ 22.4 \% & 909.15 & 99.04 & $^{93}$Nb(d,$\alpha$2n)\newline $^{89}$Nb decay\newline  & -7768.43\newline -41097.4 \\ \hline 
$^{88}$Zr\newline 0$^{+}$\newline  & 83.4 d & EC 100 \%\newline  & 392.87 & 97.29 & $^{93}$Nb(d,$\alpha$3n)\newline $^{88}$Nb decay\newline  & -17087.81 \newline -53618.2 \\ \hline 
$^{90m}$Y\newline 7$^{+}$\newline 682.04  keV & 3.19 h & IT 99.9982\%\newline $\betaup^{-}$ 0.0018 \% & 202.53\newline 479.51 & 97.3\newline 90.74 & $^{93}$Nb(d,p$\alpha$)\newline  & +2703.71 \\ \hline 
$^{87m}$Y\newline 9/2$^{+}$\newline 380.82keV & 13.37 h & IT 98.43 \%\newline ${\beta}^{+}$ 0.75 \%\newline EC 0.75 \% & 380.79 & 78.06 & $^{93}$Nb(d,?p3n)\newline $^{87}$Zr decay\newline  & -24986.77\newline -57736.59 \\ \hline 
$^{87}$Y\newline 1/2$^{-}$ & 79.8 h & EC 99.82 \%\newline ${\beta}^{+}$ 0.180 \% & 388.531\newline 484.805 & 82.2\newline 89.8 & ${}^{93}$Nb(d,$\alpha$p3n)\newline $^{87}$Zr decay & -24986.77\newline -57736.59 \\ \hline 
\end{tabular}

\end{center}
\begin{flushleft}
\tiny{\noindent The Q-values refer to formation of the ground state. In case of formation of a higher laying isomeric state it should be corrected with the level energy of the isomeric state shown in Table 1. When complex particles are emitted, instead of individual protons and neutrons, the Q-values have to be decreased by the respective binding energies of the compound particles: np-d, +2.2 MeV; 2np-t, +8.48 MeV; 2p2n-$\alpha$, +28.30 MeV.}
\end{flushleft}

\end{table*}

%\setcounter{table}{1}

%\begin{table*}[t]
%\tiny
%\caption{continued}
%\centering
%\begin{center}

%\end{center}
%\end{table*} 

\section{3.	Comparison with nuclear model calculations}
\label{3}
The cross sections of the investigated reactions were compared with the data given in the last two on-line TENDL libraries to show the development  of the predictions (\cite{15} and \cite{16}). These libraries  are based on both default and adjusted TALYS (1.6) calculations \cite{17}. The cross sections of the investigated reactions were calculated by us using ALICE-IPPE \cite{18} and EMPIRE-II [19] codes modified for deuterons by Igantyuk \cite{20, 21} . Independent data for isomers with ALICE-D code were obtained by using the isomeric ratios calculated with EMPIRE-II.

\section{Results}
\label{4}
The measured experimental cross-section data are shown in Figs. 1-10 together with the results of the earlier measurements and of the theoretical calculations. The numerical values are presented in Tables 2 - 3.

\subsection{The $^{93}$Nb(d,2n)$^{93m}$Mo reaction}
\label{4.1}
The radionuclide 93Mo has a metastable state with a half-life of 6.85 h and a long-lived (T$_{1/2}$ = 4.0 10$^3$ a) ground state. In the gamma spectra only the lines of the metastable state were detected. In the literature, apart from our earlier data \cite{3, 4}, also a recent measurement by Avrigeanu et al. \cite{6} was found and the four experimental data sets are agree well in the overlapping energy region (Fig. 1). All theoretical predictions significantly overestimate the experimental data (especially in case of TENDL-2015).

\begin{figure}
\includegraphics[scale=0.3]{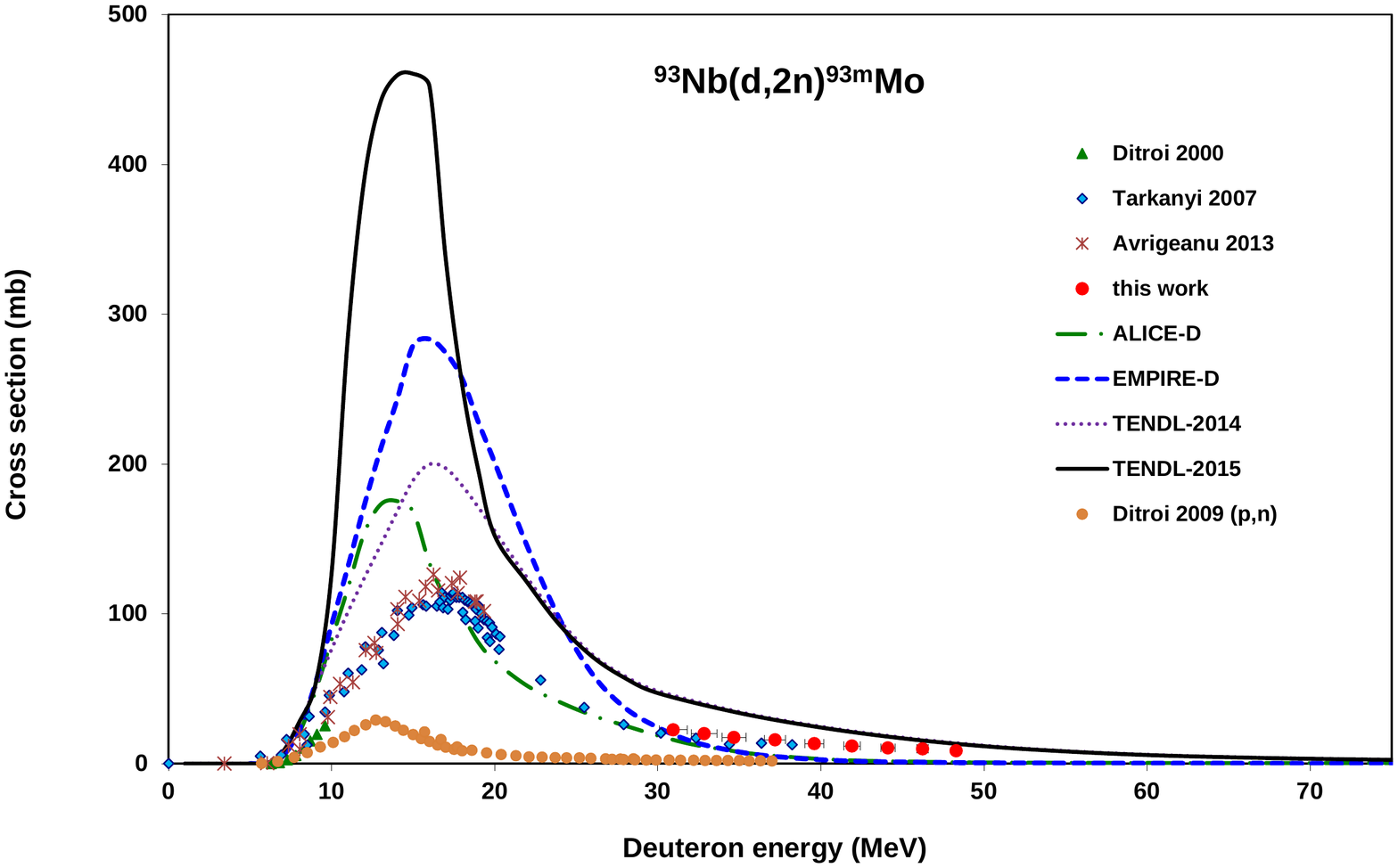}
\caption{Excitation function of the $^{93}$Nb(d,2n)$^{93m}$Mo reaction compared with the earlier results and with  the theoretical model calculations}
\end{figure}

\subsection{The $^{93}$Nb(d,5n)$^{90}$Mo reaction.}
\label{4.2}

No earlier experimental data were found in the literature. A good agreement between experimental and the theoretical results (Fig. 2) can be observed only for the TENDL libraries. Both EMPIE and ALICE strongly overestimate the experimental values.

\begin{figure}
\includegraphics[scale=0.3]{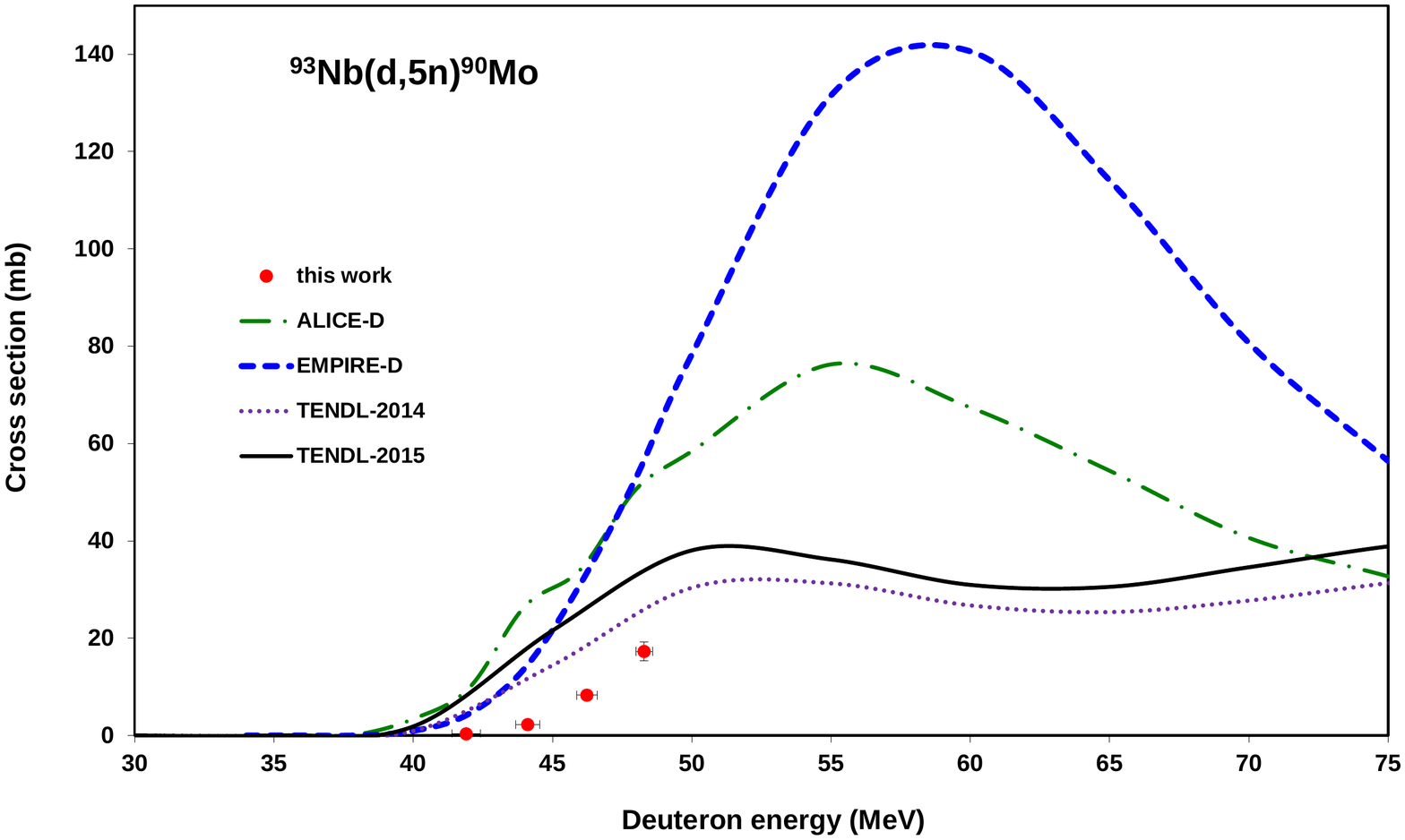}
\caption{Excitation function of the $^{93}$Nb(d,5n)$^{90}$Mo reaction compared with the earlier results and with  the theoretical model calculations}
\end{figure}

\subsection{The $^{93}$Nb(d,p2n)$^{92m}$Nb reaction.}
\label{4.3}

The only strong $\gamma$-line of the metastable (T$_{1/2}$ = 10.15 d) state of $^{92}$Nb is common with one of the -lines of the very long-lived ground state. (T$_{1/2}$ = 3.47 10$^7$ a). As this ground state can be considered quasi-stable, only cross-sections for the metastable state can be derived and are presented in Fig. 3. The agreement with our earlier measurements and the results of Avrigenau \cite{6} is good. The TENDL predictions overestimate the experimental values. The agreement with ALICE-D and Empire-D is acceptable up to 30 MeV but the behavior at higher energy is quite different.

\begin{figure}
\includegraphics[scale=0.3]{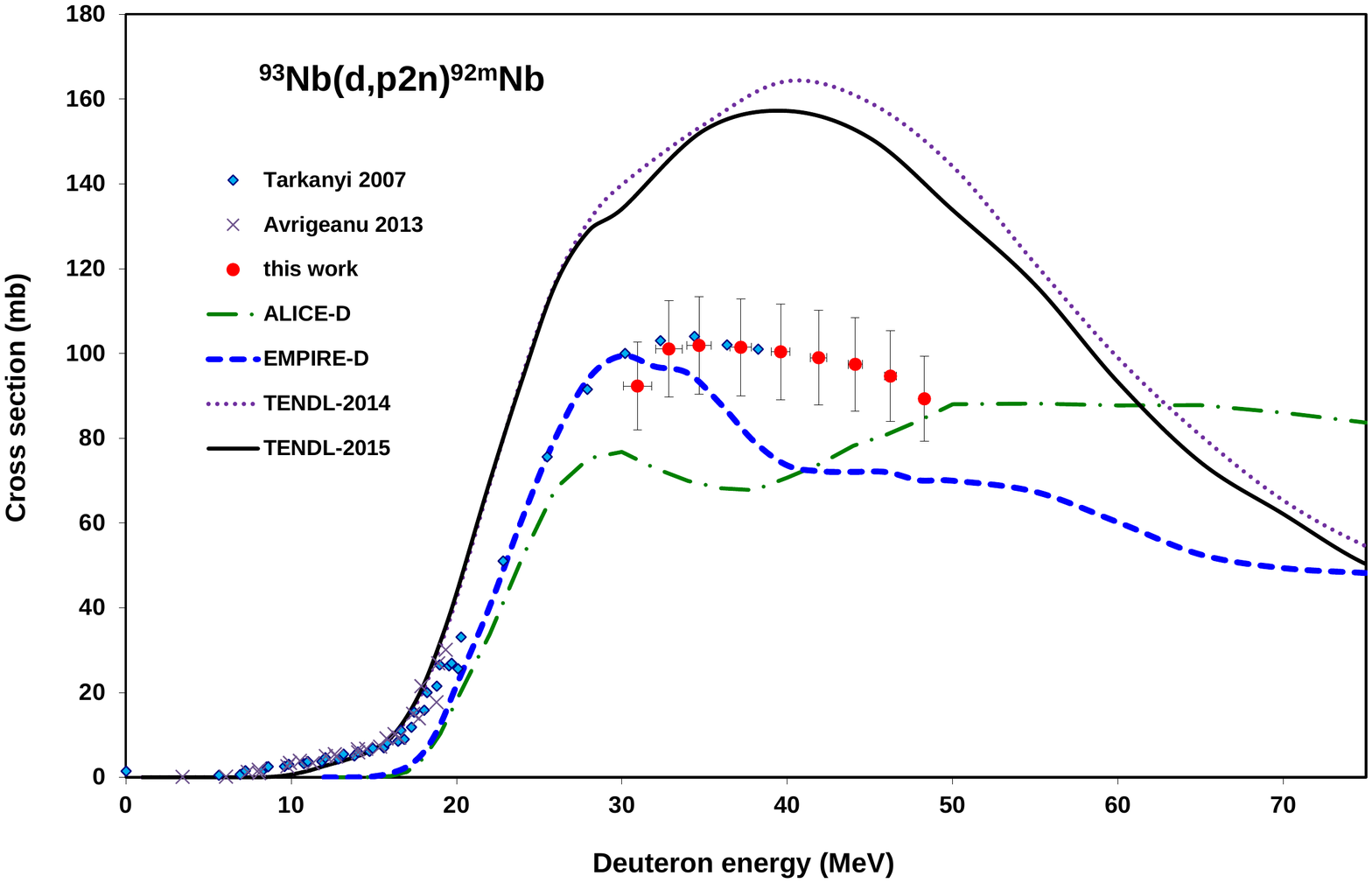}
\caption{Excitation function of the $^{93}$Nb(d,p2n)$^{92m}$Nb reaction compared with the earlier results and with  the theoretical model calculations}
\end{figure}

\subsection{The $^{93}$Nb(d,p3n)$^{91m}$Nb reaction.}
\label{4.4}

We could only measure the excitation function of the metastable $^{91m}$Nb state (T$_{1/2}$ = 60.86 d) of the $^{91}$Nb isotope. The long half-life ground state (6.8 102 a) has no gamma lines and in any case the produced activity is very low. The measured cross sections of $^{91m}$Nb are cumulative as they contain the direct formation and a contribution through decay of the shorter-lived isomeric (64.6 s, IT: 50.0\% and $\varepsilon$: 50 \%) and ground (15.49 min, $\varepsilon$: 100 \%) states of the $^{91}$Mo parent. The agreement with or earlier measurements is acceptable. The theoretical model calculations in TENDL-2014 and 2015 overestimate the experimental value with a factor of 4 (Fig. 4). The predicted values of ALICE-D and EMPIRE-D are also too high by a factor of two.

\begin{figure}
\includegraphics[scale=0.3]{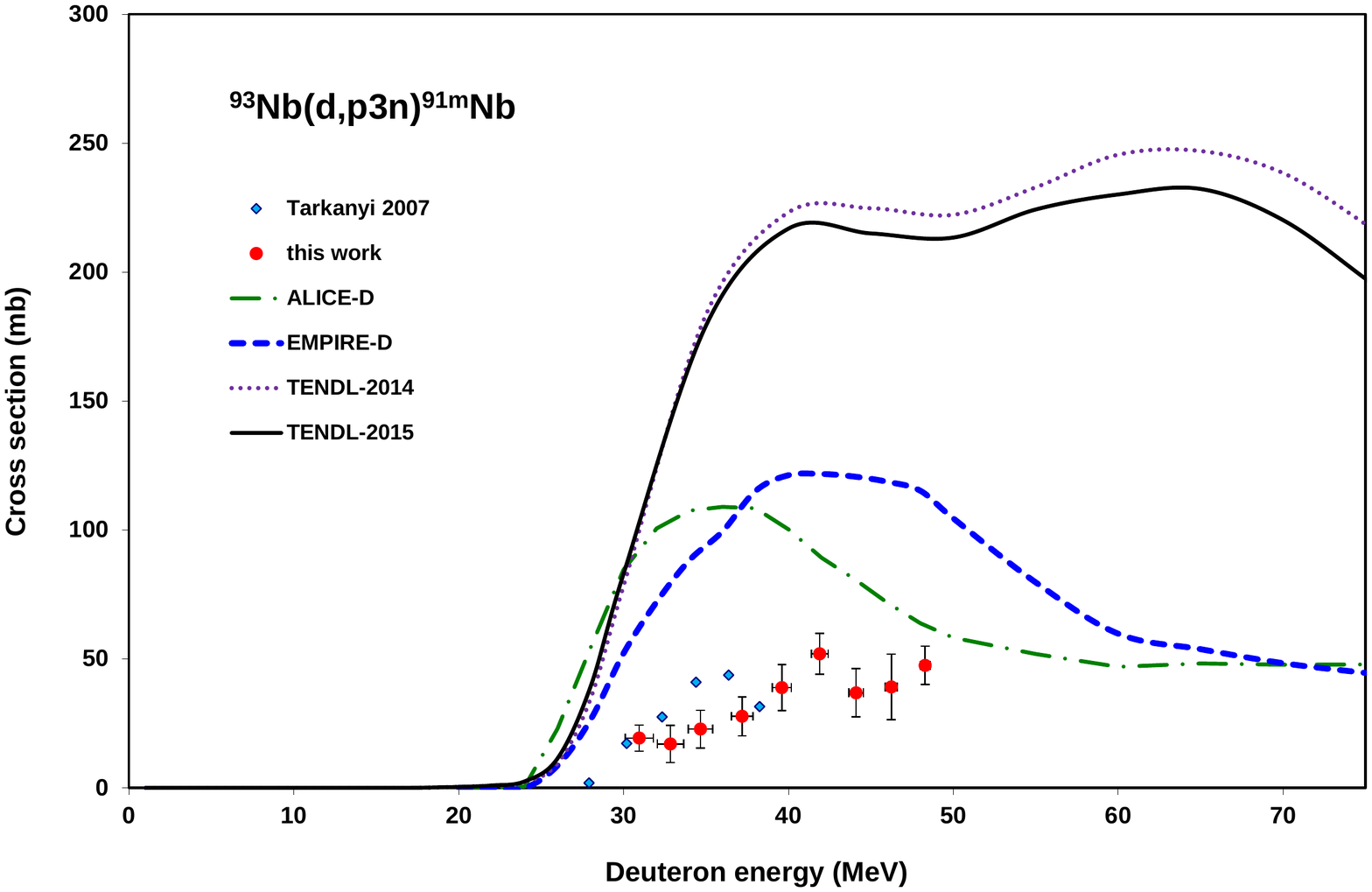}
\caption{Excitation function of the $^{93}$Nb(d,x)$^{91m}$Nb reaction compared with the earlier results and with  the theoretical model calculations}
\end{figure}

\subsection{The $^{93}$Nb(d,p4n)$^{90}$Nb reaction}
\label{4.5}
The measured activation cross sections for $^{90}$Nb (T$_{1/2}$ = 14.60 h) include the direct formation and the contribution by total decay of the short half-life isomeric state $^{90m}$Nb (18.81 s, IT: 100 \%). A good agreement was found with our earlier measurements. There are large disagreement between the TENDL data and the ALICE-D and EMPIRE-D result around the maximum (Fig. 5). Not a real improvement between the 2015 and 2014 versions of the TENDL library is seen.

\begin{figure}
\includegraphics[scale=0.3]{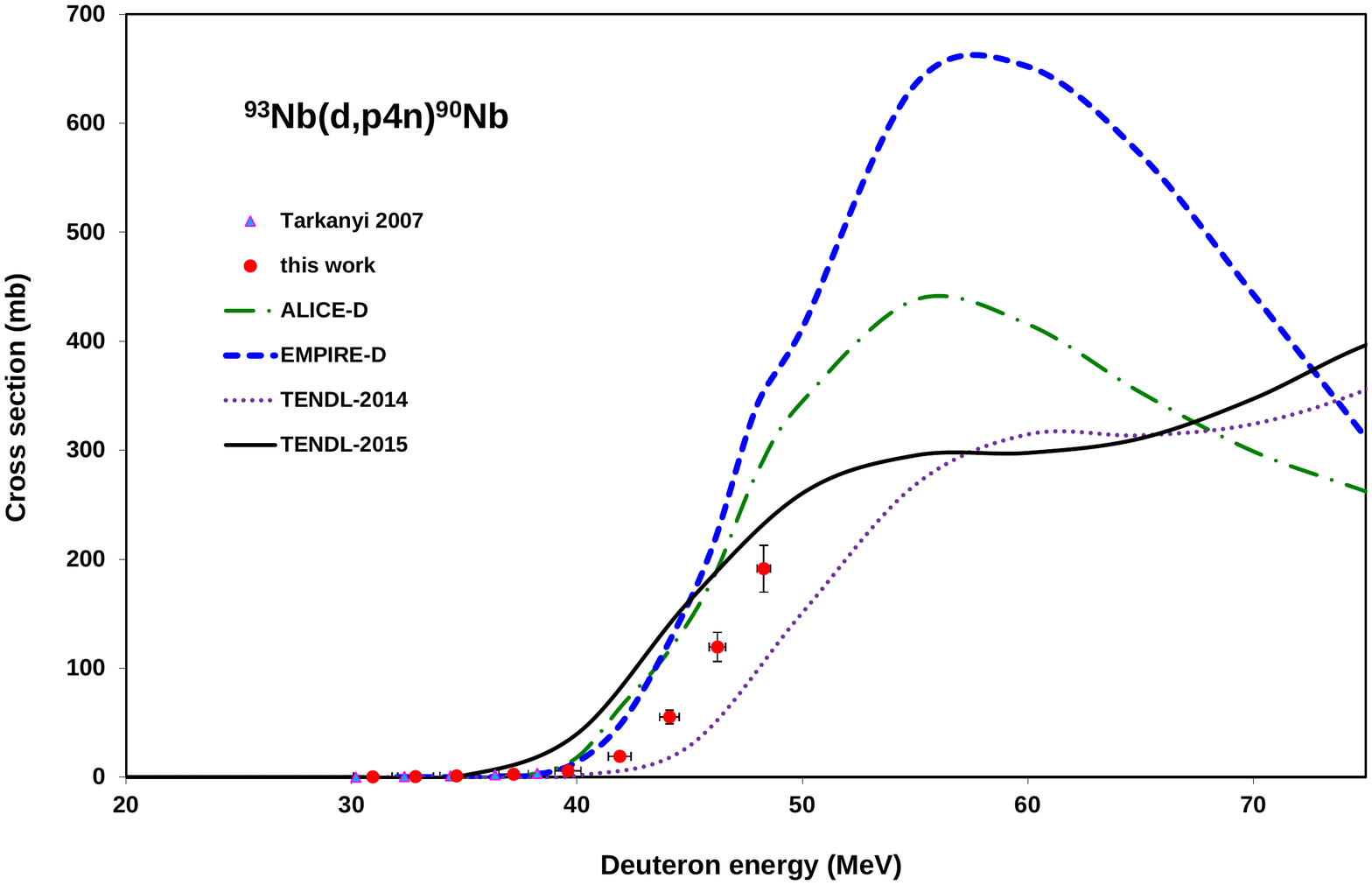}
\caption{Excitation function of the $^{93}$Nb(d,p4n)$^{90}$Nb reaction compared with the earlier results and with  the theoretical model calculations}
\end{figure}

\subsection{The $^{93}$Nb(d,x)$^{89}$Zr reaction}
\label{4.6}
The 78.9 h half-life ground state of $^{89}$Zr is formed directly via a $^{93}$Nb(d,2p4n) reaction, via decay of the directly formed short-lived metastable state $^{89m}$Zr (T$_{1/2}$ = 4.161 min,  IT: 93.77 \%) and through the  $^{89}$Mo(T$_{1/2}$ = 2.11 min, $\varepsilon$: 100 \%) $\longrightarrow$ $^{89m1}$Nb (T$_{1/2}$ = 66 min, $\varepsilon$: 100 \%) + $^{89}$Nb (T$_{1/2}$ = 2.03 h, $\varepsilon$: 100 \% ) $\longrightarrow$ $^{89m}$Zr $\longrightarrow$ $^{89}$Zr  decay chain.  The practical threshold of 20 MeV and a local maximum around 35 MeV (Fig. 6) indicate that the (d,$\alpha$2n) reaction is the only contributing pathway until  reactions with less clustered emissions play a role above 40 MeV. Our new results are in good agreement with the previous measurements. The model predictions describe more or less the overall behavior and contributions of the different pathways but there are large differences in the predicted values.  			

\begin{figure}
\includegraphics[scale=0.3]{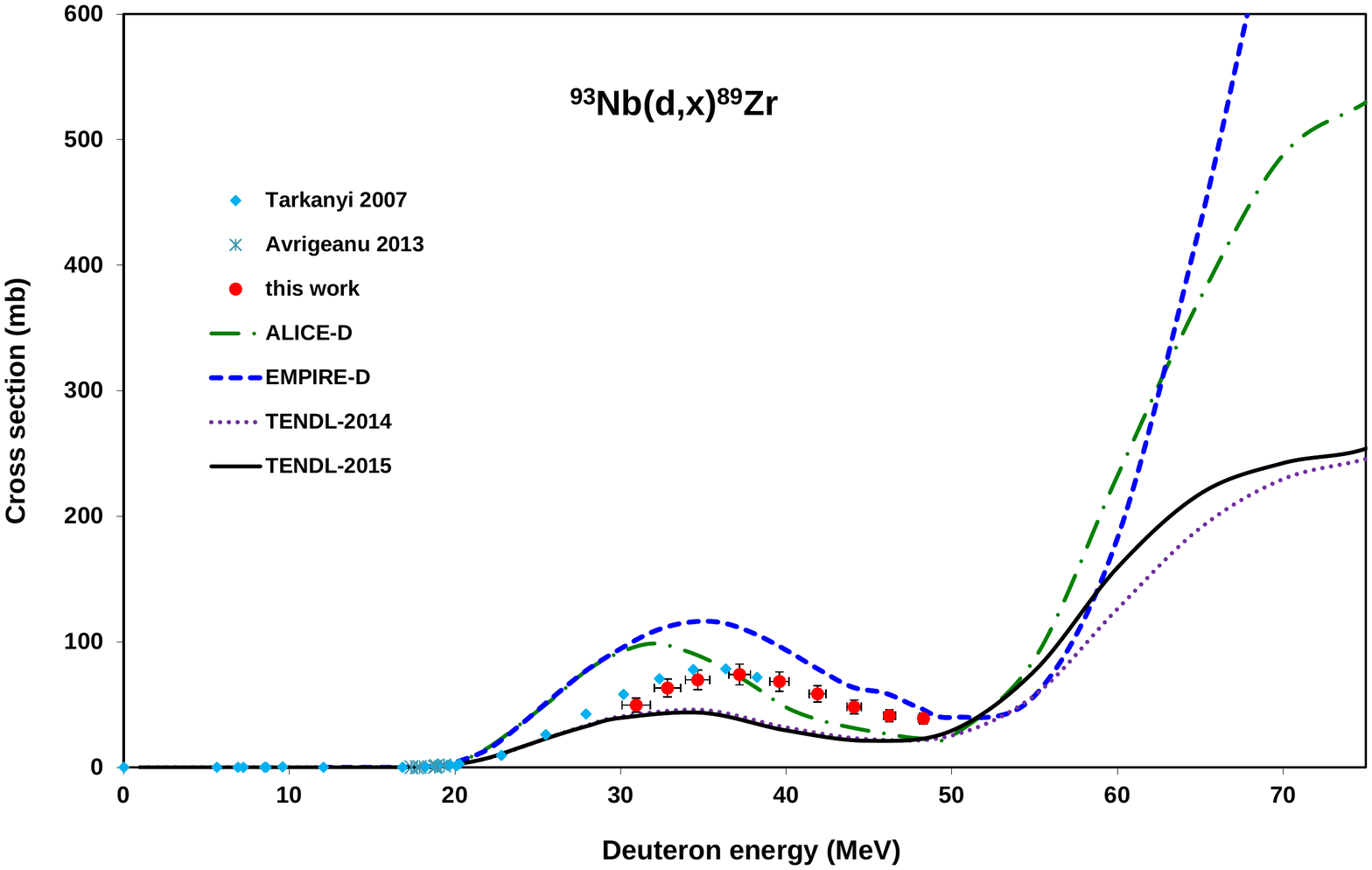}
\caption{Excitation function of the $^{93}$Nb(d,x)$^{89}$Zr reaction compared with the earlier results and with  the theoretical model calculations}
\end{figure}

\subsection{The $^{93}$Nb(d,x)$^{88}$Zr reaction}
\label{4.7}
The formation of $^{88}$Zr (T$_{1/2}$ = 83.4 d) is also cumulative: direct through the (d,2p5n) reaction and through the $^{88}$Mo (T$_{1/2}$ = 8.0 min, $\varepsilon$: 100 \%) $\longrightarrow$  $^{88m1}$Nb (T$_{1/2}$ = 7.78 min, $\varepsilon$: 100 \%) + $^{88}$Nb (T$_{1/2}$ = 14.55 min, $\varepsilon$: 100 \%) $\longrightarrow$ $^{88}$Zr decay chain.  As for the previous activation product, here too, the clustered emission is responsible for the lower energy part of the excitation function. The theoretical predictions show large disagreements (Fig. 7).

\begin{figure}
\includegraphics[scale=0.3]{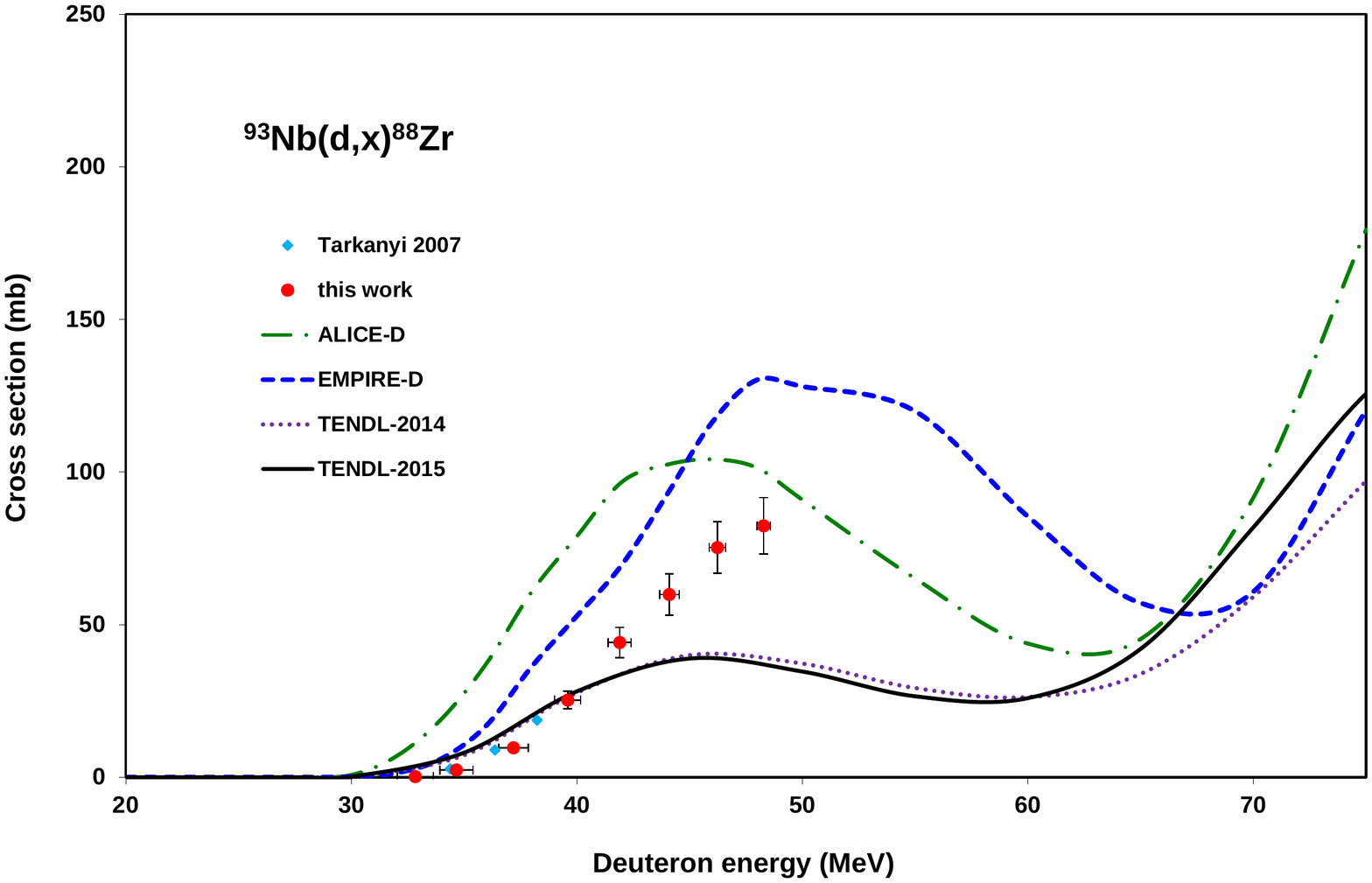}
\caption{Excitation function of the $^{93}$Nb(d,x)$^{88}$Zr reaction compared with the earlier results and with  the theoretical model calculations}
\end{figure}

\subsection{The $^{93}$Nb(d,3p2n)$^{90}$Y reaction}
\label{4.8}
The metastable state $^{90}$Y (T$_{1/2}$ = 3.19 h) is produced only directly by a high energy threshold reaction if emission of individual nucleons is considered. The possible long half-life parent $^{90}$Sr (T$_{1/2}$ = 28.79 a) decays only to the $^{90}$Y ground state. The practical experimental threshold of 10 MeV (confirmed by the theoretical predictions) shows that multiple clustered emissions play a predominant role at lower energies (Fig. 8). A good agreement between our new data and the earlier measurement is seen. According to Fig. 8 the TENDL data are significantly low, but at least the shape of excitation function is similar to the experiment. A local maximum in the EMPIRE-D and ALICE –D results is seen around 25 MeV, which is not found in the experimental data. The magnitudes of the theoretical cross sections of all codes show also large disagreement. The longer-lived ground state emits only a very weak, high energy $\gamma$-line (2186 keV, 1.4 10-6 \%) and was not detected in our experiment. 

\begin{figure}
\includegraphics[scale=0.3]{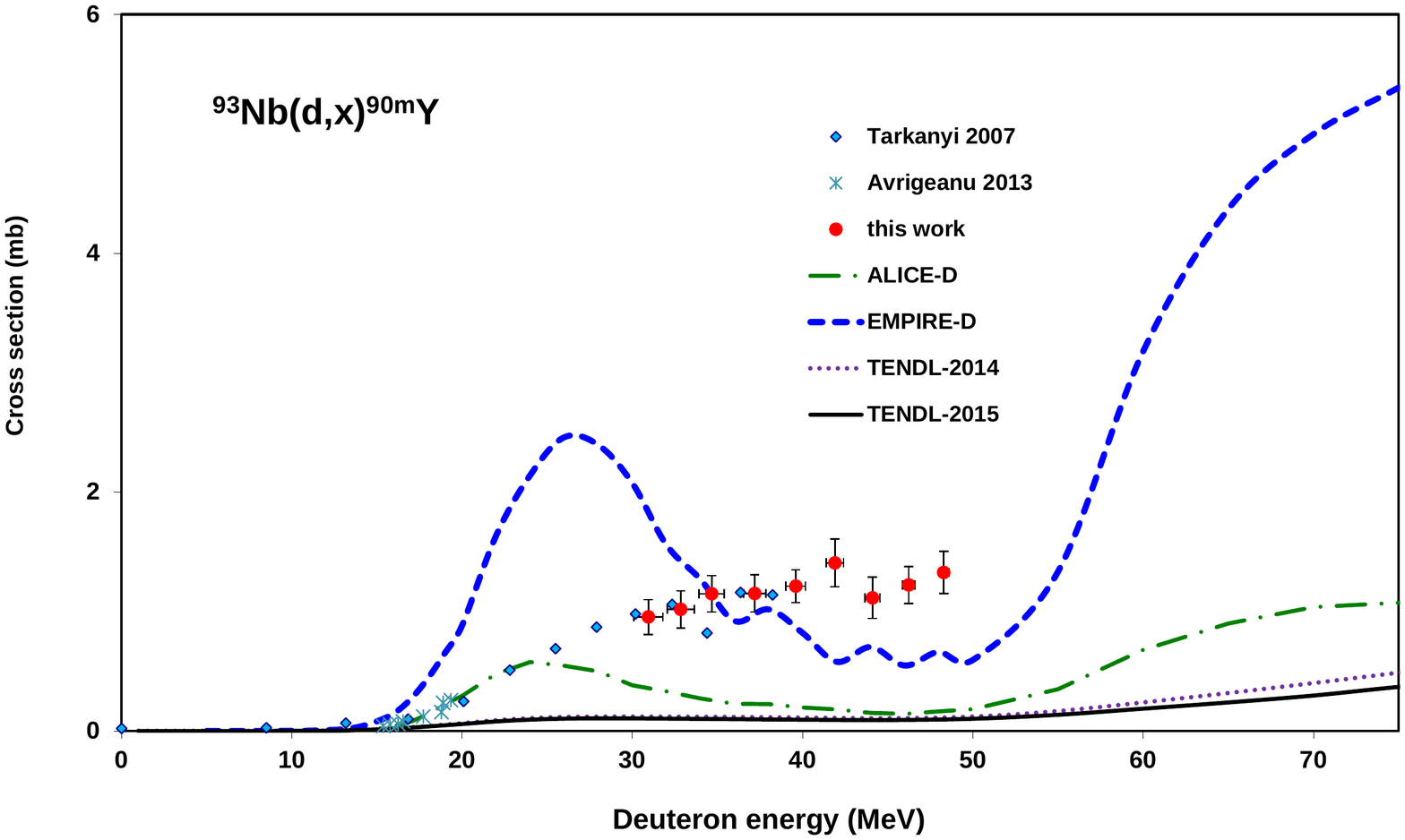}
\caption{Excitation function of the $^{93}$Nb(d,3p2n)$^{90}$Y reaction compared with the earlier results and with  the theoretical model calculations}
\end{figure}

\subsection{The $^{93}$Nb(d,x)$^{87m}$Y reaction}
\label{4.9}
Apart from the direct (d,3p5n), high threshold reaction and lower threshold multiple cluster emission, the metastable state $^{87m}$Y (T$_{1/2}$ = 13.37 h) is fed by the decay chain $^{87}$Mo(13.4 s, $\varepsilon$: 100 \%) $\longrightarrow$  ($^{87m1}$Nb (T$_{1/2}$ = 2.6 min, $\varepsilon$: 100 \% $^{87}$Nb (T$_{1/2}$ = 3.75 min , $\varepsilon$: 100 \% )) $\longrightarrow$ ($^{87m}$Zr (T$_{1/2}$ = 14.0 s,  IT: 100 \%) + $^{87}$Zr (T$_{1/2}$ = 1.6 h, $\varepsilon$: 99 \% )) $\longrightarrow$ $^{87m}$Y. The most contributing pathway is probably the $^{93}$Nb(d,$\alpha$4n)$^{87m,g}$Zr reaction that from systematics is known to have rather high cross sections. The theoretical results show large mutual disagreements (Fig. 9).

\begin{figure}
\includegraphics[scale=0.3]{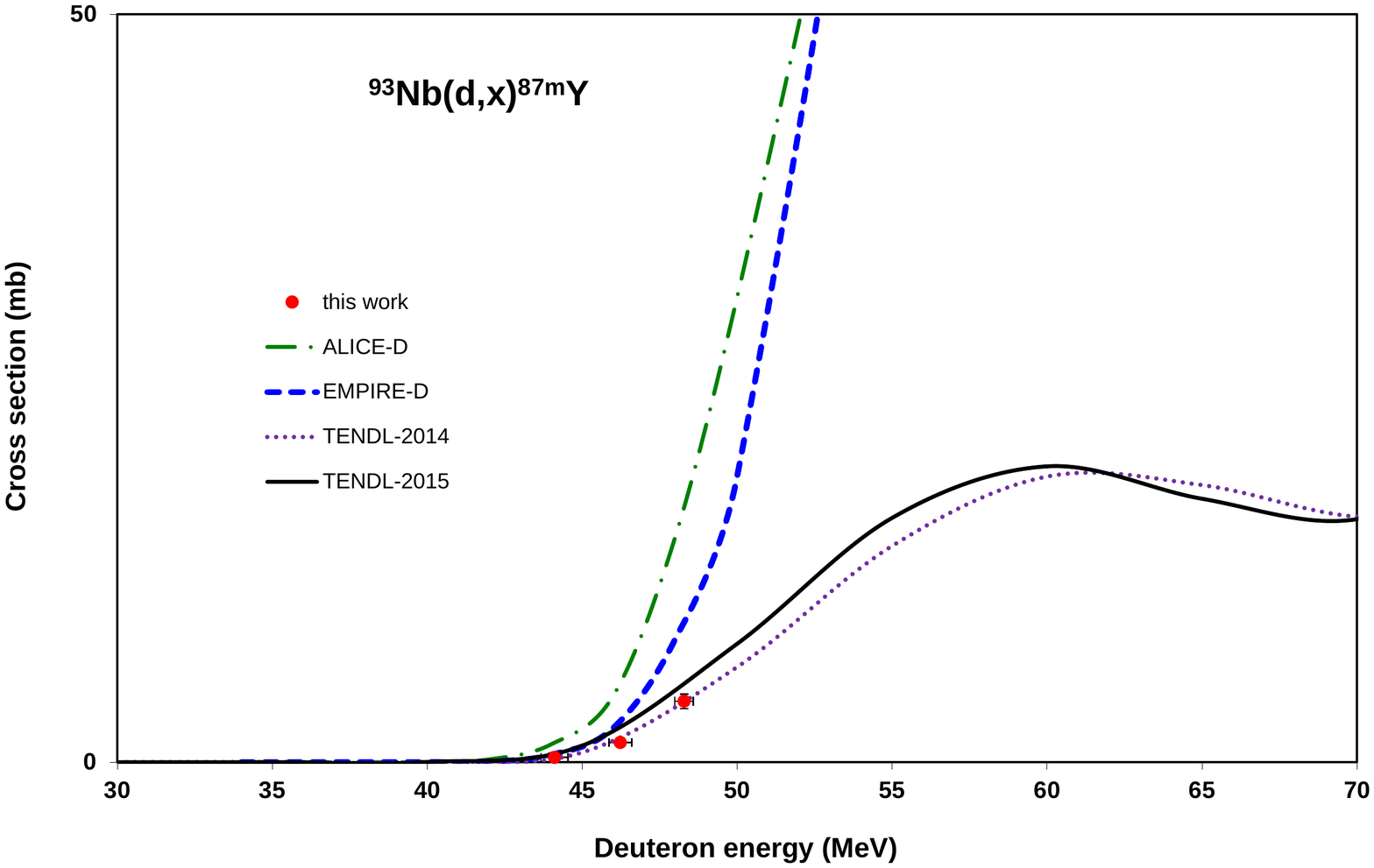}
\caption{Excitation function of the $^{93}$Nb(d,x)$^{87m}$Y reaction compared with the earlier results and with  the theoretical model calculations}
\end{figure}

\subsection{The $^{93}$Nb(d,x)$^{87g}$Y reaction}
\label{4.10}
The cumulative production cross-sections of the longer-lived ground state $^{87g}$Y (T$_{1/2}$ = 60.3 h) were measured after complete decay of the possible parents $^{87}$Mo (100 \%), $^{87}$Nb (100 \%), $^{87}$Zr (99+1 \%), $^{87m}$Y (IT: 98.43 \%) radioisotopes (see above). As we can expect that the $^{93}$Nb(d,$\alpha$4n)$^{87m,g}$Zr reaction has the highest cross-sections of all possible channels, the cross sections of the $^{87g}$Y(cum) should differ max 2.5 \% from the cross section of the $^{87m}$Y (1 \% from $^{87}$Zr direct decay  and 1,5 \% from  missing $^{87m}$Y($\varepsilon$: 1.57 \%). The present cross section data have around 12 \% total uncertainty, from which the uncertainty of the detector efficiency is around 5 \% (the cross sections of the $^{87m}$Y and the $^{87}$Y were obtained from spectra measured at different source-detector distance). Therefore the uncertainties did not allow showing the differences between excitation functions for metastable and ground sate formation.  But the Figures 9 and 10 show that the cross-sections are similar. There are large disagreements between results of different theoretical codes also for cumulative production of $^{87g}$Y. The best result is given by the TENDL-2014 again, while the TENDL-2015 does not show improvement.

\begin{figure}
\includegraphics[scale=0.3]{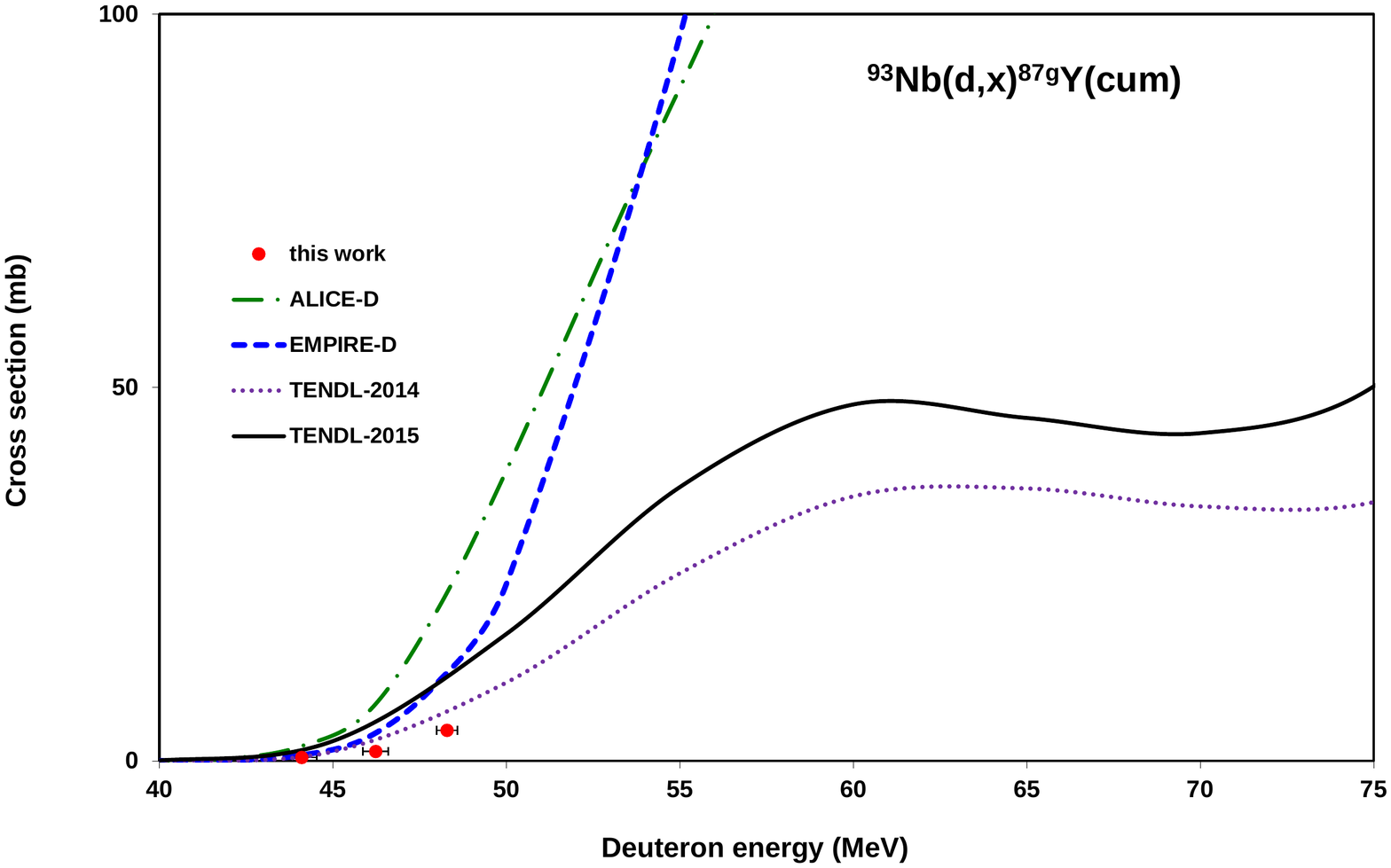}
\caption{Excitation function of the $^{93}$Nb(d,x)$^{87g}$Y reaction compared with the earlier results and with  the theoretical model calculations}
\end{figure}

\begin{table*}[t]
\tiny
\caption{Experimental cross sections of the Mo and Nb radioisotopes}
\centering
\begin{center}
\begin{tabular}{|p{0.2in}|p{0.1in}|p{0.2in}|p{0.1in}|p{0.2in}|p{0.1in}|p{0.2in}|p{0.1in}|p{0.2in}|p{0.1in}|p{0.2in}|p{0.1in}|} \hline 
\multicolumn{2}{|p{0.3in}|}{\textbf{Energy\newline (E$\pm\Delta$E)\newline (MeV)}} & \multicolumn{10}{|p{1.5in}|}{\textbf{Cross section ($\sigma\pm\Delta\sigma$)\newline (mb)}} \\ \hline 
\multicolumn{2}{|c|}{\textbf{}} & \multicolumn{2}{|c|}{\textbf{$^{93m}$Mo}} & \multicolumn{2}{|c|}{\textbf{$^{90}$Mo}} & \multicolumn{2}{c|}{\textbf{$^{92m}$Nb}} & \multicolumn{2}{|c|}{\textbf{$^{91m}$Nb}} & \multicolumn{2}{|c|}{\textbf{$^{90}$Nb}} \\ \hline 
48.3 & 0.3 & 8.59 & 0.98 & 17.3 & 1.9 & 89.3 & 10.0 & 47.5 & 7.5 & 191.3 & 21.5 \\ \hline 
46.2 & 0.4 & 9.91 & 1.12 & 8.29 & 0.94 & 94.7 & 10.7 & 39.2 & 12.7 & 119.4 & 13.4 \\ \hline 
44.1 & 0.4 & 10.5 & 1.2 & 2.24 & 0.26 & 97.4 & 11.0 & 36.9 & 9.4 & 55.2 & 6.2 \\ \hline 
41.9 & 0.5 & 11.8 & 1.3 & 0.32 & 0.05 & 99.0 & 11.1 & 52.0 & 7.9 & 19.2 & 2.2 \\ \hline 
39.6 & 0.6 & 13.3 & 1.5 & ~ &  & 100.4 & 11.3 & 38.9 & 8.9 & 5.81 & 0.66 \\ \hline 
37.2 & 0.7 & 15.9 & 1.8 & ~ &  & 101.5 & 11.4 & 27.7 & 7.5 & 2.64 & 0.31 \\ \hline 
34.7 & 0.7 & 17.5 & 2.0 & ~ &  & 101.9 & 11.5 & 22.8 & 7.3 & 1.07 & 0.15 \\ \hline 
32.8 & 0.8 & 22.5 & 2.5 & ~ &  & 101.0 & 11.4 & 17.1 & 7.2 & 0.46 & 0.07 \\ \hline 
30.9 & 0.9 & 20.0 & 2.3 & ~ &  & 92.3 & 10.4 & 19.3 & 5.1 & 0.15 & 0.02 \\ \hline 
\end{tabular}
\end{center}
\end{table*}

\begin{table*}[t]
\tiny
\caption{Experimental cross sections of the Zr and Y radioisotopes}
\centering
\begin{center}
\begin{tabular}{|p{0.2in}|p{0.1in}|p{0.2in}|p{0.1in}|p{0.2in}|p{0.1in}|p{0.2in}|p{0.1in}|p{0.2in}|p{0.1in}|p{0.2in}|p{0.1in}|} \hline 
\multicolumn{2}{|p{0.3in}|}{\textbf{Energy\newline (E$\pm\Delta$E)\newline (MeV)}} & \multicolumn{10}{|p{1.5in}|}{\textbf{Cross section ($\sigma\pm\Delta\sigma$)\newline (mb)}} \\ \hline
\multicolumn{2}{|c|}{\textbf{}} & \multicolumn{2}{|c|}{\textbf{$^{89}$Zr}} & \multicolumn{2}{|c|}{\textbf{$^{88}$Zr}} & \multicolumn{2}{|c|}{\textbf{$^{90}$${}^{m}$Y}} & \multicolumn{2}{|c|}{\textbf{$^{87m}$Y}} & \multicolumn{2}{|c|}{\textbf{$^{87g}$Y}} \\ \hline 
48.3 & 0.3 & 39.0 & 4.4 & 82.4 & 9.3 & 1.33 & 0.18 & 4.07 & 0.49 & 4.09 & 0.47 \\ \hline 
46.2 & 0.4 & 41.2 & 4.7 & 75.3 & 8.5 & 1.22 & 0.15 & 1.31 & 0.20 & 1.27 & 0.17 \\ \hline 
44.1 & 0.4 & 48.2 & 5.4 & 59.9 & 6.8 & 1.12 & 0.17 & 0.31 & 0.05 & 0.45 & 0.08 \\ \hline 
41.9 & 0.5 & 58.5 & 6.6 & 44.2 & 5.0 & 1.41 & 0.20 & ~ &  & ~ &  \\ \hline 
39.6 & 0.6 & 68.3 & 7.7 & 25.3 & 2.9 & 1.21 & 0.14 & ~ &  & ~ &  \\ \hline 
37.2 & 0.7 & 74.0 & 8.3 & 9.63 & 1.09 & 1.15 & 0.15 & ~ &  & ~ &  \\ \hline 
34.7 & 0.7 & 69.6 & 7.8 & 2.36 & 0.38 & 1.15 & 0.15 & ~ &  & ~ &  \\ \hline 
32.8 & 0.8 & 63.3 & 7.1 & 0.26 & 0.21 & 0.95 & 0.15 & ~ &  & ~ &  \\ \hline 
30.9 & 0.9 & 49.5 & 5.6 & ~ &  & 1.02 & 0.16 & ~ &  & ~ &  \\ \hline 
\end{tabular}

\end{center}
\end{table*} 

\section{Integral yields}
\label{5}
The comparison of the yields for production of $^{93m}$Mo, $^{92m}$Nb, $^{91m}$Nb, $^{90}$Nb, $^{89}$Zr, $^{88}$Zr, $^{90}$Y up to 40 MeV calculated from a fit to the excitation functions can be found in \cite{3}. The yields calculated from these earlier excitation functions completed with our new results  are shown in Figs. 11 and 12  in comparison with the earlier measured experimental yields of  Dmitriev et al. \cite{22} and Konstantinov et al. \cite{23}. Only those literature values are displayed on the figures, which are not overlapping. The calculated integral yields are so called physical yields \cite{24, 25} i.e. “yields for an instantaneous irradiation”. For the $^{87}$Y isotopes no yield curves were calculated, because there were only 3 experimental point for each reaction.

\begin{figure}
\includegraphics[scale=0.3]{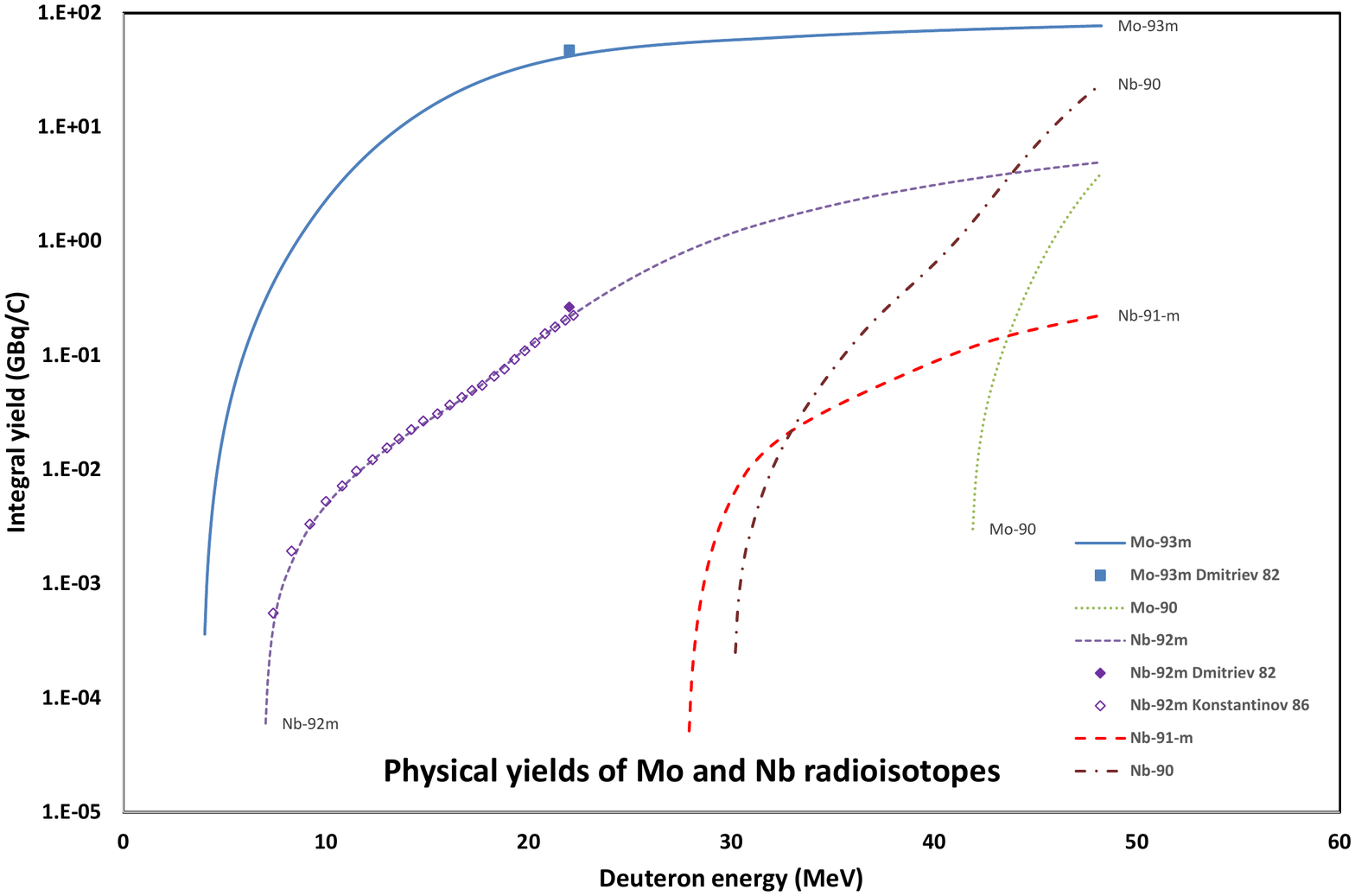}
\caption{Integral yields for production of Mo and Nb radioisotopes}
\end{figure}

\begin{figure}
\includegraphics[scale=0.3]{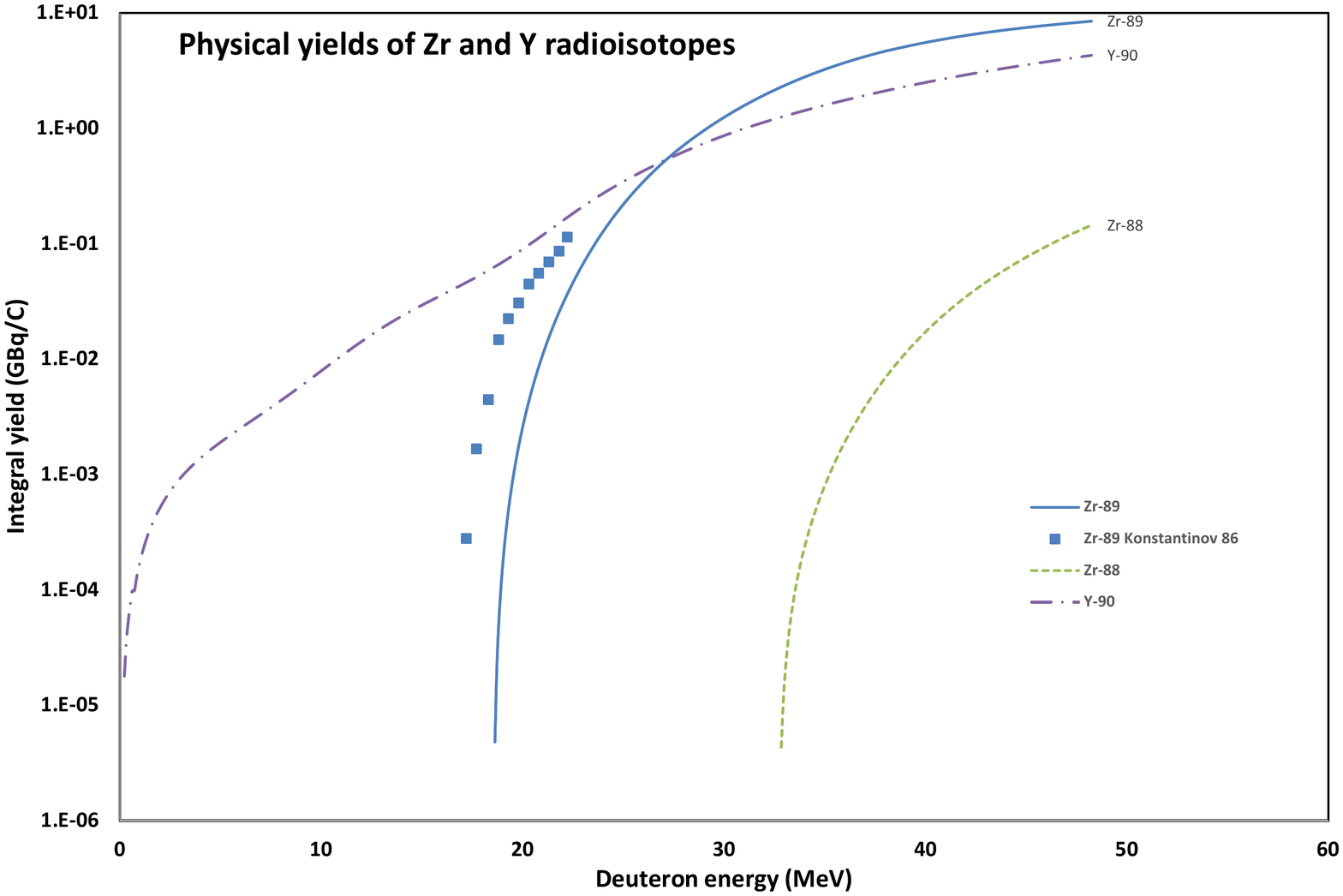}
\caption{Integral yields for production of Zr and Y radioisotopes}
\end{figure}

\section{Applications}
\label{6}

The element Nb and its alloys, due to their useful physical and chemical properties, are important materials from the point of view of applications. Activation cross section of proton and deuteron induced reactions are important for the nuclear industry,  accelerator technology \cite{26}, thin layer activation controlled wear measurements \cite{27} and  nuclear medicine. 
The measured deuteron induced activation data \cite{3} show that niobium has a very low activation up to 25 MeV energy (except for formation of the short half-life $^{93m}$Mo), while the produced activity significantly increases above 40 MeV.
We are here discussing in some detail the possible applications in the field of the medical isotope production and thin layer activation. Among the investigated reaction products a few radioisotopes are potential candidates for use in nuclear medicine, such as $^{93m}$Mo, $^{90}$Nb, $^{89}$Zr and the $^{88}$Zr/$^{88}$Y generator. We shortly review the production routes from the point of view of usefulness of deuteron induced reactions on niobium.
In this section we also discuss the industrial applicability of some produced isotopes in the field of wear, corrosion and erosion measurements (radioisotope tracing).

\subsection{$^{93m}$Mo production}
\label{6.1}
The molybdenum radionuclide $^{93m}$Mo (T$_{1/2}$ = 6.85 h), essentially decaying by IT, has a moderate half-life and high intensity gamma-lines making it suitable for diagnostic nuclear medicine \cite{28, 29}. The high energy of these lines (> 600 keV) makes this affirmation doubtful.
The main light ion induced  routes to produce nca (no-carrier added) $^{93m}$Mo are:  $^{93}$Nb(p,n)$^{93m}$Mo, $^{93}$Nb(d,2n)$^{93m}$Mo, $^{nat}$Zr($\alpha$,xn) reactions, $^{nat}$Zr($^{3}$He,xn) reaction and  the $^{nat}$Y($^7$Li,3n) \cite{28} reaction.  The cross sections of the (d,2n) and (p,n) reactions on the same monoisotopic $^{93}$Nb target are compared in Fig. 1, which shows that the deuteron induced reaction has 4 times higher cross-sections extending over a wider energy range, hence leading to higher yields.

\subsection{$^{90}$Nb production}
\label{6.2}
The radioisotope $^{90}$Nb is a positron emitter with a positron branching of 51 \% and a rather low β+-energy of Emean = 662 keV (E$_{max}$ = 1.5 MeV). Its half-life of 14.6 h makes it especially promising for the quantitative investigation through positron emission tomography (PET) of biological processes with slow distribution kinetics \cite{30, 31}.
It can be produced directly via $^{90}$Zr(p,n), $^{90}$Zr(d,2n), $^{89}$Y($\alpha$,2n) and $^{89}$Y($^{3}$He,n) reactions and through decay of the $^{90}$Mo (T$_{1/2}$ = 5.6 h) parent formed by $^{93}$Nb(p,4n) or $^{93}$Nb(d,5n) reactions. The reactions on $^{90}$Zr have high production yields, but require highly enriched targets that, however, should not be too expensive, due to the high abundance of $^{90}$Zr (51.45 \%). The bombarding beams needed for the $^{89}$Y($\alpha$,2n) and $^{89}$Y($^{3}$He,n) reactions are presently  available only at a few centers. The yields are lower and a high intensity $^{3}$He beam is expensive. The proton and deuteron induced reactions on $^{93}$Nb require high energy machines. Moreover, the irradiation time for the indirect production is limited by the half-life of $^{90}$Mo, and some stable $^{93}$Nb will be also present after decay of separated molybdenum from the decay of the simultaneously produced isomeric state $^{93m}$Mo with comparable half-life. Other contaminating Mo radioisotopes have much shorter half-life and can be eliminated by an appropriate cooling period.

\subsection{$^{89}$Zr production}
\label{6.3}
The radionuclide $^{89}$Zr (T$_{1/2}$ = 78.4 h, $\beta^+$: 22.4 \%) is used in PET. Its relatively long half-life allows  to use high-resolution PET/CT to localize and image tumors with monoclonal antibody radiopharmaceuticals and thus potentially expand the use of PET \cite{32}.
There are two interesting pathways that have been  used  for the production of $^{89}$Zr:  the $^{89}$Y(p,n)$^{89}$Zr and  $^{89}$Y(d,2n)$^{89}$Zr reactions. The (p,n) route is more competitive: smaller cyclotrons can be used (lower energy for the maximum of the excitation curve), similar cross sections \cite{33}, and higher yields due to the lower stopping. Other production routes through alpha and $^{3}$He particle induced reactions on isotopes of zirconium and high energy proton and deuteron induced reactions on $^{93}$Nb are also possible. The $^{3}$He and alpha particle beams are only available at a few centers. The yields are low and highly enriched targets are required. The yields of the higher energy proton and deuteron induced reactions on $^{93}$Nb are in principle not so low if thick targets could be used. However the incident energy is limited by the possible production of contaminating long-lived $^{88}$Zr.

\subsection{$^{88}$Y production}
\label{6.4}
The radionuclide $^{90}$Y (T$_{1/2}$ = 64.0 h) is used in radio-immunotherapy. It has a relatively short half-life and low abundant $\gamma$-lines (2186.242 keV, 1.4x 10$^{-6}$ \%) but emits low energy X-rays and conversion electrons. The radioisotope $^{88}$Y has much longer half-life (T$_{1/2}$ = 106.627 d) and high intensity gamma-rays hence, by substituting it for $^{90}$Y, could give the possibility to follow radiopharmaceutical development of $^{90}$Y labeled products \cite{34}. Large scale production of $^{88}$Y is presently mostly done via the spallation reaction on Mo with high energy protons. There are alternative, low energy, direct and indirect methods (through $^{88}$Zr (T$_{1/2}$ = 83.4 d) decay) for his production.
Proton and deuteron irradiations on Mo, Nb, Zr and Y were investigated by us \cite{35}. As at that time no experimental data were available for $^{93}$Nb(d,x)$^{88}$Zr and $^{93}$Nb(d,x)$^{88}$Y, these reactions were  not included in the yield comparison published in the above reference (partly obtained from theoretical results), from which the following conclusions were drawn: 
\begin{itemize}
\item The direct production of $^{88}$Y is low compared to $^{88}$Zr formation for both particles on Mo, Nb, Zr and Y targets, except for deuteron-induced reactions on Y targets.
\item Significant amounts of $^{88}$Zr can be produced at low energy accelerators by using Y and Zr targets.
\item At higher energy accelerators Nb targets result in higher production rates compared to Mo, but the yield is still lower than the yield on Zr and Y targets at lower energies. 
\end{itemize}
Some other important alternative direct production routes also exist and were included in our comparison although no experimental data of our group were available. It concerned  the $^{nat}$Sr(p,xn)$^{88}$Y, $^{nat}$Sr(d,xn)$^{88}$Y,natRb(α,xn)$^{88}$Y and natRb($^{3}$He,xn)$^{88}$Y reactions, that were however investigated by other authors. It can be concluded that direct production of $^{88}$Y with $^{nat}$Sr(p,xn)$^{88}$Y, $^{nat}$Sr(d,xn)$^{88}$Y reactions can be done with high yield.
The results presented here for production of $^{88}$Y and $^{88}$Zr do not change the earlier conclusions. The proton and deuteron induced reactions on niobium can only be a taken into account as a satellite or by-product production method (for example production in the target holders made from niobium used in production of long-lived 69Ge from gallium targets \cite{36, 37}).

\subsection{Thin layer activation}
\label{6.5}
Because niobium is an important construction and alloying material in nuclear industry and in other industrial fields, it is worth to study the possibility of TLA (Thin Layer Activation) by using the produced isotopes. There are several criteria concerning the half-life, $\gamma$-radiation, production yield of the isotopes in order to classify them as proper isotope for thin layer activation (wear, corrosion and erosion measurement). Among the isotopes studied in this work the $^{92m}$Nb (T$_{1/2}$ = 10.15 d, E$_{\gamma}$ = 934.44 keV, I$_{\gamma}$ = 99.15 \%), $^{91m}$Nb (T$_{1/2}$ = 60.86 d, E$_{\gamma}$ = 1204.67 keV, I$_{\gamma}$ = 2.0 \%), $^{88}$Zr (T$_{1/2}$ = 83.4 d, E$_{\gamma}$ = 271.8 keV, I$_{\gamma}$ = 33.2 \%, E$_{\gamma}$ = 671.2 keV, I$_{\gamma}$ = 66.5 \%, E$_{\gamma}$ = 1057.01 keV, I$_{\gamma}$ = 99.95 \%, E$_{\gamma}$ = 1082.53 keV, I$_{\gamma}$ = 99.95 \%) fulfill the criteria. $^{92m}$Nb has the highest cross section (yield) and the $\gamma$-energy is suitable, but the half-life is a bit too short making possible to investigate relatively quick processes. $^{91m}$Nb and $^{88}$Zr have the most suitable half-lives, the $^{91m}$Nb has proper but weak $\gamma$-line and the $^{88}$Zr has a set of proper $\gamma$-radiations. Both $^{91m}$Nb and $^{88}$Zr have lower production cross sections. In order to compare their radioactive tracing capabilities some of the depth profile curves are shown in Fig. 13.
The optimum irradiation energy for $^{92m}$Nb is 36.1 MeV in order to achieve homogeneous activity distribution under the surface. The homogeneity criterion is fulfilled (within 1 \% accuracy) in this case down to 73 $\mu$m depth by 15$^o$ irradiation angle and 2 days waiting time after the irradiation. In the case of perpendicular irradiation this depth will be four times larger. In the case of $^{91m}$Nb the optimum energy is 42.5 MeV for quasi-homogeneous distribution, in this case the depth of homogeneity is only 21 $\mu$m with the same irradiation conditions (except the bombarding energy) (see Fig. 13). Because by $^{88}$Zr we did not reach the local maximum of the excitation function in our measurement, and this maximum is expected to be around 50 MeV, we have chosen the maximum experimental energy for this example, i.e. 46.47 MeV. From Fig. 13 it is seen, that in this case only a linear distribution can be produced, while the homogeneous distribution requires higher bombarding energy. The linearity is fulfilled down to a depth of 150 $\mu$m. It is obviously seen that in spite of its shorter half-life the $^{92m}$Nb is the most proper candidate for wear studies.

\begin{figure}
\includegraphics[scale=0.3]{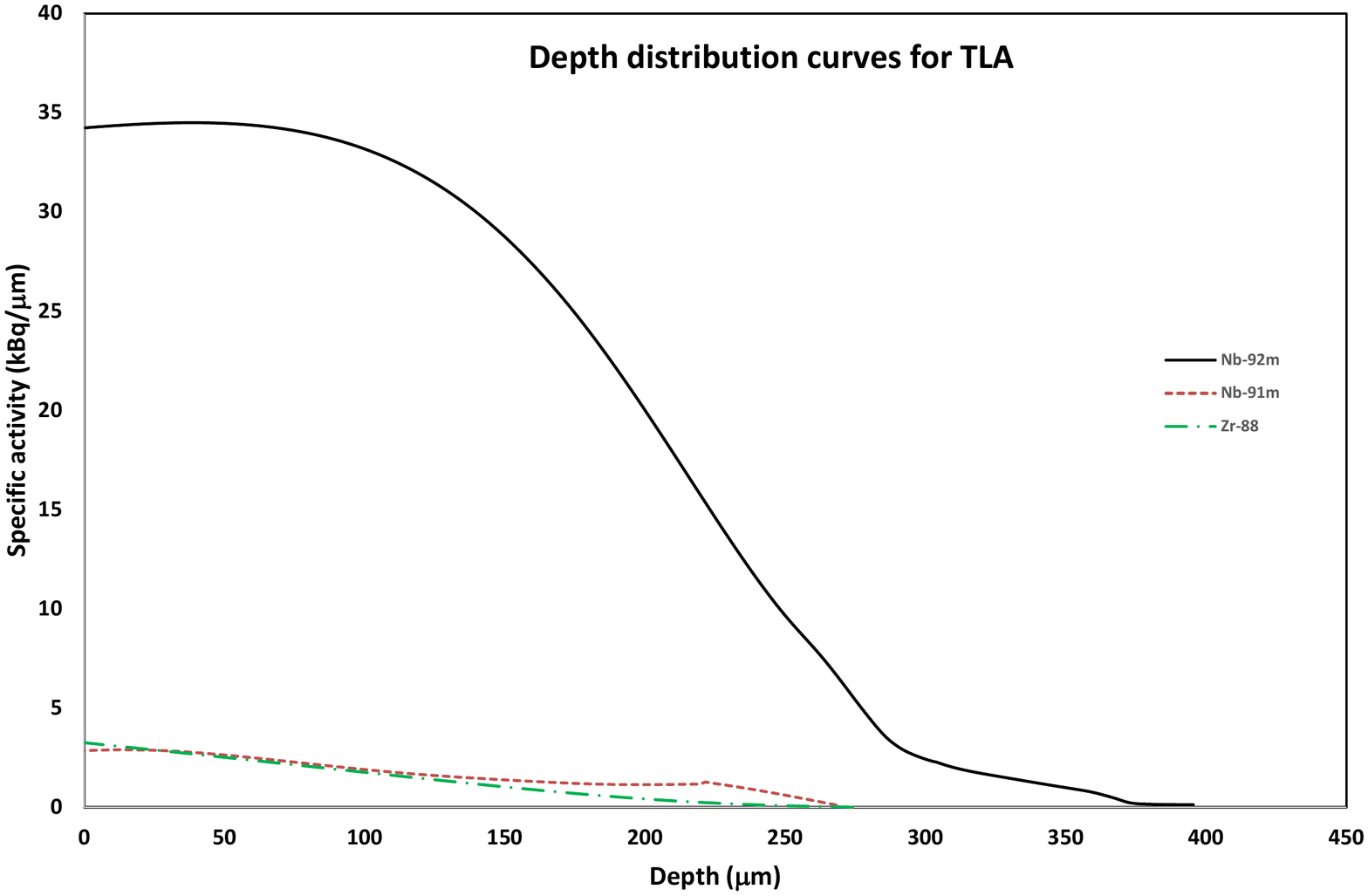}
\caption{Depth profile curves for TLA after 2 days waiting time for 1 hour 1 $\mu$A irradiation under 15$^o$. The bombarding energies are 36.1, 42.5 and 46.47 MeV for $^{92m}$Nb, $^{91m}$Nb and $^{88}$Zr respectively}
\end{figure}

\section{Summary}
\label{7}
We present experimental excitation functions for $^{93}$Nb(d,x)$^{93m,90}$Mo, $^{92m,91m,90}$Nb, $^{89,88}$Zr and $^{88,87m,87g}$Y formation in the energy range 30–50 MeV. All data above 37 MeV deuteron energy are first experimental results. The results were compared with the theoretical cross-sections and it is shown that the description for formation of the investigated radioproducts is poor. For medically relevant activation products the deuteron induced reactions on niobium have a perspective for production of $^{93m}$Mo and $^{90}$Nb. The  investigated higher energy activation data can be applied, apart for studies of  the nuclear reaction mechanism and medical isotope production, in other important fields like activation analysis, nuclear astrophysics, space applications(resistance of electronics, shielding, etc.), accelerator technology (safeguards, shielding, beam monitoring, targetry), tracing of industrial and biological processes (thin layer activation, special radioisotopes, nanoparticles). Three from the produced isotopes were also investigated from the point of view of Thin Layer Activation and it has been proved that all the three are suitable for performing wear measurements under different requirements. The most proper is the $^{92m}$Nb with the drawback of its shorter half-life.

\section{Acknowledgements}

This work was performed in the frame of the HAS-FWO Vlaanderen (Hungary-Belgium) project. The authors acknowledge the support of the research project and of the respective institutions (VUB, LLN) in providing the beam time and experimental facilities.
%\FloatBarrier
 


%% References

%\clearpage
\bibliographystyle{elsarticle-num}
\bibliography{Nbd}






% \end{thebibliography}

\end{document}